\documentclass[12pt]{article}
\large

\title{A class of exactly solvable quantum models of scalar gravity}

\vspace{0.5cm}

\author{ V.V. Belokurov$^{1,2}$ and E.T. Shavgulidze$^{1}$    \\
\\    {\em 1. Lomonosov Moscow State University, Russia }
\\    {\em 2. Institute for Nuclear Research of Russian Academy of Sciences, Russia }
\\ {\it  belokur@rector.msu.ru ; shavgulidze@bk.ru}}

\date{ \ \ \  }
\begin{document}
\maketitle

\begin{abstract}

Exactly solvable quantum theory of a singular at the origin scalar field  with the self-interaction of Liouville type is proposed.
The mean value of the scale factor in the FLRW metric as a function of conformal time is  evaluated explicitly. 

\end{abstract}

\vspace{2cm}

In our recent papers \cite{(BSh1)}, \cite{(BSh2)} we have studied the quantum field models with functional integrals being defined over the space of singular functions. Here, we propose a class of exactly solvable quantum models of scalar gravity and using the method of the previous papers give the explicit solution of the quantum problem.

For the last three decades low dimensional dilaton gravity has been caused a great interest in various
contexts.\footnote{For a review see e.g. \cite{(GKV)}.}

Due to the results of \cite{(Pol)} and \cite{(D)}, \cite{(DK)} the quantization of 2-D gravity is reduced to the quantization of the Liouville field theory (see also \cite{(P)}, \cite{(H)} and \cite{(M)}).

On the other hand, the most popular approach to explain the evolution of the universe is based on the interaction of gravity with a scalar field. Various models of self-interacting scalar fields have being used for this purpose starting with the pioneer papers  on inflation \cite{(S)}, \cite{(G)}, \cite{(L)} and up to the recent ones (see, e.g.\cite{(OR)}, \cite{(vH)}).

 In this paper we consider a model of quantum scalar field with the classical action
\begin{equation}
   \label{1}
 \tilde{A}(\varphi )=\frac{1}{2}\int \limits _{0}^{T}\left \{ (\dot{\varphi}(t))^{2}dt - \alpha \lambda \, e^{\alpha \varphi (t)} + \lambda ^{2}\, e^{2\alpha \varphi (t)}\right \}dt 
\end{equation}
 of a generic Liouville type. However, for the model to be exactly solvable the coefficients at different terms are chosen in a special way. Moreover, as we will see later on, to evaluate the functional integrals explicitly we need that $\alpha < 0 $ and $\lambda < 0\,.$

Bearing in mind that the scale factor in conformal coordinates for the FLRW metric is determined by the scalar field
\begin{equation}
   \label{2}
g(t)\equiv a^{2}(t)=e^{2\varphi (t)}\,,
\end{equation}
we impose the boundary condition corresponding to initial singularity
\begin{equation}
   \label{3}
g(0)=0\,,\ \ \ \ \varphi(0)=-\infty\,.
\end{equation}

In just the same way as we do it in \cite{(BSh1)}, \cite{(BSh2)}, by the substitution
\begin{equation}
   \label{4}
\xi(t)=\varphi(t) - \varphi (0)+ \lambda\,\int \limits _{0}^{t}e^{\alpha \varphi (\tau)}d\tau
\end{equation}
we turn the quantum theory defined by the action $\tilde{A}(\varphi)$ into that of the free field\footnote{As usual, the normalizing factors at functional integrals are set equal to 1.}
$$
\int\limits _{X}\,F(\varphi)\,\exp\left\{ - \tilde{A}(\varphi) \right\}\ d\varphi \,=
$$
\begin{equation}
   \label{5}
\int\limits _{C[0,1]}\,F(\varphi (\xi))\exp\left\{ -\frac{1}{2}\int \limits _{0}^{T}(\dot{\xi}(t))^{2}dt \right\}\ d\xi\equiv \int\limits _{C[0,1]}\,F(\varphi (\xi))\,w(d\xi)\,.
\end{equation}
Here, $X$ is the space of functions that have a singularity of the form $\varphi\sim -\frac{1}{\alpha}\ln t $ at the origin $t=0\,.$ And eq. (\ref{4}) is understood as the limit of a regularized equation. 

Actually, eq. (\ref{5}) can be considered as the definition of the functional measure on the space $X$ (see \cite{(BSh2)}).

Note that the second term in the action $\tilde{A}(\varphi)$ appears in left-hand side of eq. (\ref{5}) as the result of the Ito integral
$$
\int \limits _{t=0}^{t=T}\,\lambda \, e^{\alpha \varphi (t)}\,d\varphi\,.
$$
Here we have omitted the boundary terms as they are unessential in the functional integral (are canceled by the normalizing factor).

The validity of eq. (\ref{5}) can be easily verified using the discrete version of Wiener integrals. And for an invertible substitution
\begin{equation}
   \label{6}
\xi(t)=\varphi(t)+ \int \limits _{0}^{t}f( \varphi (\tau))d\tau
\end{equation}
the equality
$$
\int\,F(\xi)\exp\left\{ -\frac{1}{2}\int \limits _{0}^{1}(\dot{\xi}(t))^{2}dt \right\}\ d\xi\,=
$$
\begin{equation}
   \label{7}
\int\,F(\xi(\varphi))\,\exp\left\{ -\frac{1}{2}\int \limits _{0}^{T}\left((\dot{\varphi}(t))^{2}+f^{2}(\varphi(t))-f'(\varphi(t))\right)dt\,+\,b.t.   \right\}\ d\varphi
\end{equation}
can be proven as well.

The equality of the functional integrals (\ref{5}) leads to the effect deduced in \cite{(BSh2)}. The classical action
$\tilde{A}(\varphi) $ and the classical limit $A(\varphi)$ of the corresponding quantum theory turn out to be different.

The classical limit of the right-hand side of eq. (\ref{5}) gives the equation
\begin{equation}
   \label{8}
\ddot{\xi}(t)=0\,.
\end{equation}
Taking into account the boundary condition
$$
\dot{\xi}(T)=0\,,
$$
we get the classical equation of motion
\begin{equation}
   \label{9}
\dot{\xi}(t)=0\,.
\end{equation}
Eq. (\ref{8}) is equivalent to the  equation
\begin{equation}
   \label{10}
\ddot{\varphi}(t))-\alpha \lambda^{2}e^{2\alpha\varphi(t)}=0\,,
\end{equation}
that is nothing else than the Euler-Lagrange equation of the action
\begin{equation}
   \label{11}
 A(\varphi )=\frac{1}{2}\int \limits _{0}^{T}\left \{ (\dot{\varphi}(t))^{2}dt  + \lambda ^{2}\, e^{2\alpha \varphi (t)}\right \}dt \,,
\end{equation}
but not that of the action $\tilde{A}(\varphi) $ ( eq. (\ref{1}) ).

The true classical field $ \varphi_{c}(t)$ is, therefore, the solution of eq. (10)
\begin{equation}
   \label{12}
\varphi_{c}(t)=-\frac{1}{\alpha}\ln (\alpha \lambda t)\,,\ \ \ \alpha <0,\ \lambda <0\,.
\end{equation}
Note that $\xi_{c}(t)=0$ as it follows from eq. (9) and the initial condition $\xi(0)=0\,.$

 If we inverse eq. (4) we get the explicit form of the function $\varphi(\xi)$
\begin{equation}
   \label{13}
\varphi(t)=\xi(t)-\frac{1}{\alpha}\ln\left(\alpha \lambda \int \limits _{0}^{t}e^{\alpha \xi (\tau)}d\tau \right)\,.
\end{equation}
Thus the scale factor $a(t)$ in terms of $\xi(t)$ looks like
\begin{equation}
   \label{14}
a(t)=e^{\varphi(t)}=e^{\xi(t)}\left(\alpha \lambda \int \limits _{0}^{t}e^{\alpha \xi (\tau)}d\tau \right)^{-\frac{1}{\alpha}}\,.
\end{equation}

 Now, in the quantum theory given by the action $\tilde{A}(\varphi) $ we can evaluate the mean value of the scale factor, moments and other quantities connected with it.  
 
 For the models with
$$
\alpha =-\frac{1}{n}\,, \ \ \ \ n=1,2,\ldots\ ,
$$
the evaluation can be performed explicitly (Wiener integrals are reduced to iterated Gaussian integrals).

Let e.g. $\alpha=-1\,.$ The mean value of the scale factor is
\begin{equation}
   \label{15}
<a(t)>=-\lambda\,\int\limits _{C[0,1]}\,e^{\xi(t)}\,\int \limits _{0}^{t}e^{\alpha \xi (\tau)}d\tau\,w(d\xi)=
(-2\lambda)\left(e^{\frac{t}{2}}-1 \right)\,.
\end{equation}
And the dispersion is
$$\Delta^{2}_{a}(t)=<a^{2}(t)>-<a(t)>^{2}=
$$
\begin{equation}
   \label{16}
\lambda^{2}\int\limits _{C[0,1]}e^{2\xi(t)}\left(\int \limits _{0}^{t}e^{\alpha \xi (\tau)}d\tau\right)^{2}w(d\xi)-4\lambda^{2}\left(e^{\frac{t}{2}}-1 \right)^{2}=\frac{1}{3}\lambda^{2}\left(2e^{2t}-12e^{t}+16e^{\frac{t}{2}}-6 \right).
\end{equation}

At small $t$ the mean value and the dispersion are
\begin{equation}
   \label{17}
<a(t)>=-\lambda \,t  \ \ \ (\lambda <0)\,, \ \ \ \ \ \Delta^{2}_{a}(t)=\frac{1}{3}\lambda^{2}\,t^{3}\,.
\end{equation}
As $\Delta_{a}\sim t^{\frac{3}{2}}\ll t$ at $t$ small enough, there is a region where the mean value $<a(t)>$ can be considered as the exact value of the scale factor.

Note that singularities are used to cause obstacles in many conventional classical theories. It is due to the singularity the classical action $\tilde{A}$ is inappropriate to describe the dynamics of the universe.

In our approach the existence of singularities is essential for the adequate formulation of quantum theory. In this way we can take into account  quantum effects that not only describe the dynamics at small $t$ but modify the classical theory (changing $\tilde{A}$ to $A$).


\begin{thebibliography}{14}

\bibitem {(BSh1)} V.V. Belokurov  and E.T. Shavgulidze, Paths with singularities in functional integrals of quantum
field theory. arXiv:1112.3899v2. [hep-th].

\bibitem {(BSh2)} V.V. Belokurov  and E.T. Shavgulidze, Quantum restoration of broken symmetries. arXiv:1303.3523v1. [math-ph].

\bibitem {(GKV)} D. Grumiller, W. Kummer, D.V. Vassilevich Phys. Rept. 369 (2002), 327, (arXiv:hep-th/0204253v9 Jan 2008).

\bibitem {(Pol)} A.M. Polyakov, Phys. Lett. B103 (1981), 207.

\bibitem {(D)} F. David, Mod. Phys. Lett. A3 (1988), 651.

\bibitem {(DK)} J. Distler and H. Kawai, Nucl. Phys. B321 (1989), 509.

\bibitem {(P)} J. Polchinski, Nucl. Phys. B324 (1989), 123.

\bibitem {(H)} E. D'Hoker, Mod. Phys. Lett. A6 (1991), 745.

\bibitem {(M)} R.B. Mann, Nucl. Phys. B418 (1994), 231.

\bibitem {(S)} A.A. Starobinsky, Phys. Lett. B91 (1980), 99.

\bibitem {(G)} A.H. Guth, Phys. Rev. D23 (1981), 347.

\bibitem {(L)} A.D. Linde, Phys. Lett. B108, 389 (1982), Phys. Lett. B129 (1983), 177.

\bibitem {(OR)} M. Osipov, V. Rubakov, Galileon bounce after ekpirotic contraction. arXiv:1303.1221v1  [hep-th].

\bibitem {(vH)} J.W. van Holten, On single scalar field cosmology. arXiv:1301.1174v2 [gr-qc].

\end{thebibliography}
\end{document}